\documentclass[10pt]{article}
\usepackage[letterpaper]{geometry}
\usepackage{hicss}
\usepackage{times}
\usepackage[none]{hyphenat}
\usepackage{url}
\usepackage{latexsym}
\usepackage{indentfirst}
\usepackage{graphicx}
\graphicspath{{images/}}
\usepackage[
    style=apa,
  ]{biblatex}
\pdfoutput=1
\usepackage{hyperref}
\hypersetup{breaklinks=true, colorlinks,allcolors=blue}

\usepackage{tablefootnote}
\usepackage[table,xcdraw]{xcolor}

\DeclareFieldFormat{doi}{\space\href{https://doi.org/#1}{#1}} 

\title{A Survey-Based Quantitative Analysis of Stress Factors and Their Impacts Among Cybersecurity Professionals\thanks{\,\,\,This research was supported in part by a grant under the Graduate Research Initiative from Dakota State University.}}

 \author{Sunil Arora \\
  Dakota State University \\
  {\underline{ sunil.arora@trojans.dsu.edu}} \\ \\
 \And
  John D. Hastings \\
  Dakota State University \\
  {\underline{ john.hastings@dsu.edu} } \\ \\
  }

\date{}

\begin{document}
\maketitle

\begin{abstract}
This study investigates the prevalence and underlying causes of work-related stress and burnout among cybersecurity professionals using a quantitative survey approach guided by the Job Demands-Resources model. Analysis of responses from 50 cybersecurity practitioners reveals an alarming reality: 44\% report experiencing severe work-related stress and burnout, while an additional 28\% are uncertain about their condition. The demanding nature of cybersecurity roles, unrealistic expectations, and unsupportive organizational cultures emerge as primary factors fueling this crisis. Notably, 66\% of respondents perceive cybersecurity jobs as more stressful than other IT positions, with 84\% facing additional challenges due to the pandemic and recent high-profile breaches. The study finds that most cybersecurity experts are reluctant to report their struggles to management, perpetuating a cycle of silence and neglect. To address this critical issue, the paper recommends that organizations foster supportive work environments, implement mindfulness programs, and address systemic challenges. By prioritizing the mental health of cybersecurity professionals, organizations can cultivate a more resilient and effective workforce to protect against an ever-evolving threat landscape.
\end{abstract}

\subsubsection*{Keywords:}

cybersecurity, work stress, burnout, cybersecurity professionals, mental health

\section{Introduction}\label{intro}
In today's rapidly evolving technology landscape, cybersecurity has become a necessary aspect of organizational operations, protecting critical assets ranging from hardware and software to sensitive data \parencite{cisa2023}. These assets face relentless cyber threats, which perpetually seek vulnerabilities within government agencies, public institutions, and private enterprises \parencite{admass2023cyber}. 

As cyber threats grow in sophistication and frequency, the role of cybersecurity professionals has never been more vital or more demanding \parencite{oltsik2020survey}. These individuals act as the frontline defenders against incursions, ensuring the safety and integrity of information systems. However, the immense pressure and relentless nature of their work often come at a significant cost to their mental health and well-being \parencite{tines2022}.

Despite their critical role, the mental health challenges faced by cybersecurity professionals remain underexplored and insufficiently addressed by research \parencite{singh2023stress, nobles2022stress}. The demanding responsibilities, coupled with the high stakes and constant vigilance required in cybersecurity roles, create an environment ripe for stress and burnout. A recent study \parencite{Sekuro} underscores the severity of these issues: an overwhelming majority of cybersecurity professionals reported experiencing mental health issues, and a significant number left their roles due to these formidable challenges.

Understanding the mental health challenges in cybersecurity is not only crucial for the well-being of individual professionals but also for the overall security of organizations. High levels of stress and burnout can impair decision-making \parencite{ceschi}, reduce productivity, and increase the risk of security breaches. Therefore, addressing these issues is imperative for maintaining a resilient and effective cybersecurity workforce.

This research seeks to delve deeper into the complex relationship between cybersecurity work and mental well-being. By examining the underlying causes of stress and burnout among these professionals, the study aims to illuminate the significant impacts on their mental and physical health. Furthermore, it explores the coping mechanisms employed by cybersecurity practitioners and offers insights into how organizations can better support their workforce.

To achieve these objectives, a quantitative descriptive research methodology, guided by the Job Demands-Resources (JD-R) model~\parencite{demerouti2001job}, was employed. The study used an online survey to gather data from cybersecurity professionals. The survey focused on various aspects of their work environment, stress levels, burnout experiences, and mental health challenges. By systematically analyzing the collected data, the study provides a comprehensive understanding of the factors contributing to stress and burnout in the cybersecurity field.

This study's findings aim to inform stakeholders, including organizational leaders, policymakers, and society at large, about the urgent need for targeted interventions that promote mental health. Further, the study contributes to the theoretical understanding of occupational stress and burnout by providing a view of the JD-R model's applicability in this domain.

The paper is organized as follows: Section \ref{background} provides a foundational background. Section \ref{questions} presents the research questions that guide the study. Section \ref{methodology} details the methodology, followed by the survey results in Section \ref{results}. Section \ref{discussion} discusses the results and presents recommendations. Section \ref{futurework} touches on limitations and future %research 
possibilities, and Section \ref{conclusion} concludes the paper.

\section{Background}\label{background}
Work-related stress is a mental or emotional condition caused by job demands or pressures that exceed an employee's knowledge and abilities, making it difficult for them to cope. This happens when job demands are greater than the resources and abilities of the individual \parencite{WHO2020}. Job demands include various aspects that require ongoing expertise and effort, such as workload, work pressure, complexity, and monotony. These needs span physical, mental, emotional, and social dimensions.

The resources and skills available to employees significantly impact their ability to manage job demands. Personal, organizational, and social elements can all be considered job resources and abilities.  These resources and abilities include proficiency and expertise, control over work, support networks, and a positive organizational culture \parencite{WHO2020}.

Burnout is a severe and pervasive condition that arises from prolonged workplace stress that individuals struggle to manage effectively \parencite{WHOB}. Its hallmark symptoms include emotional detachment, exhaustion, and feelings of inadequacy. Burnout can lead to physical and mental health issues, decreased productivity, and increased absenteeism \parencite{WHOB}. 

To promote well-being among cybersecurity professionals, it's crucial to understand the interplay between work-related stress, job demands, resources, and burnout. Cybersecurity professionals operate in a stressful environment that can lead to anxiety and exhaustion. The findings of a cybersecurity first responders survey detailed in \textcite{hollis} %Tammie as part of her dissertation have
revealed that job demands, including varying demands, job environments, and job content, significantly contribute to stress and mental health issues among first responders. In another study, \textcite{spears} found that jobs related to incident response in cybersecurity are associated with high levels of job stress. In a similar study \parencite{sundaramurthy}, persistently high rates of burnout experienced by security analysts in Security Operations Centers (SOCs) lead to sub-optimal decision-making when analyzing security events.

While early-career stress often comes from workload demands, seasoned professionals face additional stressors unique to their field. These challenges include perceived obstacles and misunderstandings, such as being perceived as obstacles or naysayers by business and technology staff and leaders. Another challenge is a lack of understanding from others about the intricacies of cybersecurity work. Cybersecurity experts have a weighty responsibility as they grapple with high-stakes decision-making. They believe their perspective is correct, and ignoring their advice could have detrimental consequences for the organization \parencite{patton2021navigating}. 

Despite mounting evidence indicating that cybersecurity professionals face unhealthy stress levels and burnout, scientific research in this domain remains scarce \parencite{nobles2022stress, dykstra2018cyber, singh2023stress, Sekuro}. As a result, the mental health needs of cybersecurity practitioners are often overlooked. The challenging work environment compounds these stressors, fueled by constant fears of cyberattacks, an ever-evolving threat landscape, mounting responsibilities, new regulations, and industry mandates. To address this critical gap, our research aims to identify the pivotal factors impacting cybersecurity professionals' stress levels, burnout, and overall mental health. 

According to a study by \textcite{osltiandlundell}, 71\% of cybersecurity professionals often experience unhealthy job stress levels. Another survey found that 90\% of Chief Information Security Officers (CISOs) would accept a pay cut for improved work-life balance \parencite{Sheridan, reeves2023ciso}. According to another study conducted by \textcite{nepal}, a significant proportion of cybersecurity incident responders experienced burnout due to factors, such as the demands of their jobs poor sleep quality, and the unpredictable timing of security incidents. 

Due to the ever-changing cybersecurity threats and a severe shortage of skilled professionals, organizations face the dual challenge of addressing cybersecurity threats while keeping their employees safe from burnout and exhaustion \parencite{dykstra2018cyber}. Burnout is pervasive in the field and can cause organizational resilience to suffer. Despite these concerns, organizations rely on technological solutions, neglecting the human element and creating vulnerabilities that cybercriminals can exploit \parencite{nobles2022stress}. \textcite{obi} highlighted that burnout in cybersecurity professionals negatively impacts their performance.

To improve cybersecurity defenses, initiatives that address stress and burnout must be prioritized, recognizing that the well-being of professionals impacts security outcomes. Ignoring cybersecurity stress and burnout can lead to increased data breaches, cyber-attacks, ransomware incidents, and other security breaches. Therefore, organizations must invest in their cybersecurity professionals' mental health and well-being to safeguard against potential threats \parencite{Nobles2019}.

While for-profit and non-profit entities have published numerous studies and reports, there is still a lack of academic research on this topic. A recent systematic literature review by \textcite{singh2023stress} highlights the need for a rigorous investigation into cybersecurity professionals' stress. Given the dynamic business landscape and evolving cybersecurity threat landscape, this research is becoming even more critical. Understanding and mitigating stress and burnout among cybersecurity professionals is crucial for individual resilience and organizational security. Robust academic research can provide effective strategies to address these challenges and improve the industry's overall well-being. 

\section{Research Questions}\label{questions}

This study investigates the impact of cybersecurity roles on stress, burnout, and mental health among professionals in the field. The following research questions guide the study:

\begin{enumerate}
    \item[\textbf{RQ1}] How do cybersecurity jobs and responsibilities cause high stress levels, burnout, and mental health issues in cybersecurity professionals?

    \item[\textbf{RQ2}] What are the consequences of high stress levels and burnout on cybersecurity professionals?

    \item[\textbf{RQ3}] How do cybersecurity professionals deal with high stress levels and burnout?

    \item[\textbf{RQ4}] How do these mental challenges affect the cybersecurity community?

    \item[\textbf{RQ5}] How can cybersecurity leaders and management help their cybersecurity workforce improve mental health and reduce stress?

    \item[\textbf{RQ6}] How does organizational culture affect cybersecurity professionals' stress levels and mental health?

    \item[\textbf{RQ7}] What can business and executive leaders do to help the cybersecurity department better manage their staff's mental health?
\end{enumerate}

\section{Methodology}\label{methodology}

This study employs a quantitative descriptive research methodology~\parencite{creswell2018research}, guided by the JD-R model which provides a comprehensive framework for understanding how the interaction between job demands and job resources influences employee well-being. By applying this model, the study aims to identify patterns and relationships between these factors and the levels of stress and burnout reported by professionals.

\subsection{Data Collection}

Data was collected using a survey instrument structured as follows: 

\subsubsection{Survey Questions}

Key elements of the survey questions included:
\begin{enumerate}

\item \textit{Age Restriction}: The questionnaire included a disclaimer specifying a minimum age requirement of 18 years to ensure compliance with ethical research standards. Participants were informed of this requirement in the survey introduction. Although no additional age verification was conducted, this disclaimer aimed to deter participation by minors.

\item \textit{Study Participants}:
Because the study focused on professionals in the cybersecurity domain, an initial question ensured that participants were actively employed or had worked in cybersecurity roles to qualify for participation.

\item \textit{Demographics and Contextual Information}: Questions were included to capture participants' job context, such as organization size, position level, and region of work. This data helps categorize responses and analyze how different work environments might influence stress and burnout.

\item \textit{JD-R Model Variables}:
The survey included questions about work demands (e.g., off-hours calls, long work hours, unrealistic management expectations, and the nature of cybersecurity work) and job resources (e.g., management support, well-being initiatives, tools for support, and opportunities for time off or vacations). Capturing these variables is necessary to apply the JD-R model effectively to understand the balance between demands, resources, and burnout.

\item \textit{Scope and Clarity}: Clear definitions were provided for terms such as work stress, burnout, and mental health to ensure a common understanding among participants, thereby reducing ambiguity and enhancing the reliability of responses.

\item \textit{Avoidance of Response Fatigue}: The survey was comprised of 22 questions with Yes/No options or a carefully limited set of multiple-choice answers. The brevity was designed to consider participants' time and cognitive load, thereby reducing the risk of response fatigue. The survey's conciseness encouraged higher completion rates and more engaged responses, which aimed to enhance the quality of the data collected.

\item \textit{Avoiding Bias and Leading Questions}: The questions were carefully worded to avoid leading questions that might bias responses. Questions were neutrally worded, and participants were not steered toward specific responses. This design consideration was crucial for gathering honest and accurate data.

\end{enumerate}

\subsubsection{Recruitment}

The following participant recruitment methods were utilized:

\begin{enumerate}
\item \textit{LinkedIn Connections}: Cybersecurity professionals within our LinkedIn network were directly invited to participate in the study. Leveraging existing connections facilitated targeted recruitment.

\item \textit{LinkedIn Survey Posting}: A survey link was posted on LinkedIn, inviting cybersecurity professionals to participate and respond to the questionnaire. This broader outreach aimed to engage professionals beyond immediate connections.

\item \textit{Ph.D. Cohort Peers}: Additionally, Ph.D. cohort peers who were working in the industry at the time of the study were invited to participate.
\end{enumerate}
The chosen data collection methods align with the study's objectives, emphasizing rigor and practicality.

\subsubsection{
Survey Key Features}

The questionnaire incorporated a number of key features designed to maximize success:

\begin{enumerate}

\item \textit{Online Delivery}: This study employed an online questionnaire as the primary data collection method. This approach offered a number of advantages in gathering data:
\begin{itemize}
    \item \textit{Efficiency}: Online access allowed rapidly reaching a large number of participants.
    \item \textit{Cost-Effectiveness}: Minimized costs were required compared to in-person surveys.
    \item \textit{Global Reach}: The survey was available to a global audience.
\end{itemize}

\item \textit{Accessibility and Convenience}: Participants had several weeks to access and complete the online questionnaire at their convenience. This flexibility of timing encouraged participation.

\item \textit{Anonymity and Confidentiality}: The questionnaire ensured complete anonymity, with no personally identifiable information (PII) collected. Participants could thus candidly share their experiences without privacy concerns.

\end{enumerate}

\subsection{Ethical Consideration}

The survey included an informed consent statement outlining the study’s purpose and participants' rights, including the option to withdraw at any time. The study received Institutional Review Board approval under Dakota State University IRB No. 20231129,
confirming compliance with ethical standards.

\subsection{Assumption \& Scope}

In this study, we assumed that burnout and mental health issues are present among cybersecurity professionals, as supported by prior research (see Sections \ref{intro} and \ref{background}). This assumption was essential to frame our investigation into the specific factors that contribute to these mental health challenges. By acknowledging the existence of these issues upfront, the study could focus on identifying potential causes and mechanisms behind stress and burnout rather than establishing their presence. 

Additionally, we recognized that many professionals suffering from mental illnesses keep their struggles private. The research scope was limited to cybersecurity professionals currently employed in the industry. However, the study's findings may broadly apply to other IT or organizational departments. Our commitment to secure and confidential data handling underscored our ethical approach to understanding and addressing mental health within the cybersecurity community.

\section{Results}\label{results}
The survey was available from Jan. 1 through Apr. 30, 2024, and produced the following results:

\subsection{Participant Demographics}

Our initial target was to collect responses from a minimum of 35 participants to ensure a reasonable sample size. We received 50 completed questionnaires from cybersecurity professionals across various countries (see Table \ref{table:country_comparison}) and organizations of diverse sizes (see Fig. \ref{Figure 2}).

\begin{table}[h]
\centering
\caption{Countries of work of the respondents}
\label{table:country_comparison}
\resizebox{0.45\columnwidth}{!}{%
\begin{tabular}{|l|c|}
\hline
Country & Percentage \\ \hline
United States &                    62\% \\ \hline
Canada        &                    22\% \\ \hline
Europe        &                     8\% \\ \hline
India         &                     4\% \\ \hline
Australia     &                     4\% \\ \hline
\end{tabular}
}
\end{table}

\subsection{Organizational Context}
\begin{figure}[ht!]
    \centering
    \includegraphics[width=1\linewidth]{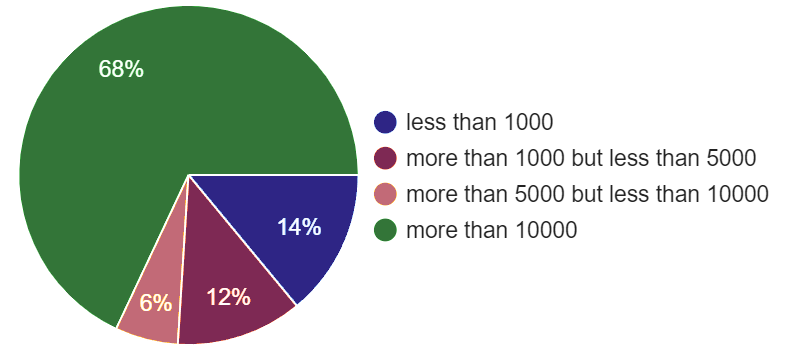}
    \caption{Organization sizes (by number of employees) of the respondents}
    \label{Figure 2}
\end{figure}

Most of the survey participants (68\%) were employed in large organizations with over 10,000 employees (see Fig. \ref{Figure 2}). 

Our participant pool represented professionals at different career levels, ranging from junior cybersecurity practitioners to senior leaders and C-suite executives who self reported their positions as seen in Table \ref{table:position_comparison}.

\begin{table}[h]
\centering
\caption{Position level of the respondents}
\label{table:position_comparison}
\resizebox{\columnwidth}{!}{%
\begin{tabular}{|l|c|}
\hline
Position & Percentage \\ \hline
Senior-level (Professional)                        &  40\% \\ \hline
Leader (Team lead/supervisor, Manager, Senior Manager) &  28\% \\ \hline
Mid-level (Professional)                           &  18\% \\ \hline
Executive (VP, SVP, 'C' class executive)           &  12\% \\ \hline
Junior level (Professional)                         & 2\% \\  \hline
\end{tabular}
}
\end{table}

\subsection{Stress Perception }

\begin{figure}[ht!]
    \centering
    \includegraphics[width=1\linewidth]{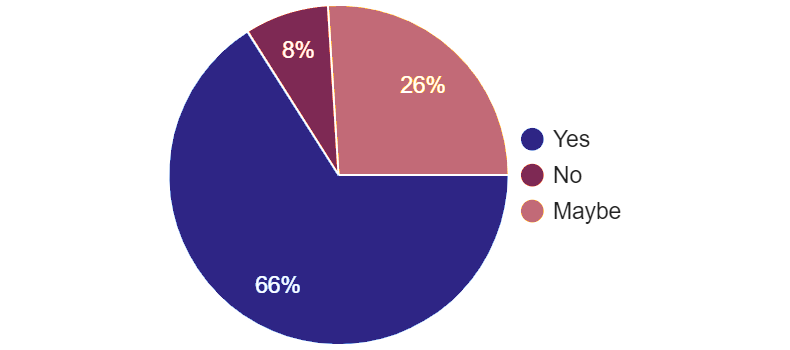}
    \caption{Respondents who feel cybersecurity jobs are more demanding}
    \label{Figure 4}
\end{figure}

A significant majority of participants expressed their belief that cybersecurity jobs are inherently more stressful than other information technology (IT) roles: 66\% of respondents indicated that cybersecurity jobs impose greater demands and stress compared to other IT positions (see Fig. \ref{Figure 4}). Given that many cybersecurity professionals start their careers in IT roles or work closely with those in IT, it seems likely that they are aware of the differences. However, the question was intended simply to gather their perceptions of the differences in stress, rather than establish it as a fact.

\begin{figure}[ht!]
    \centering
    \includegraphics[width=1\linewidth]{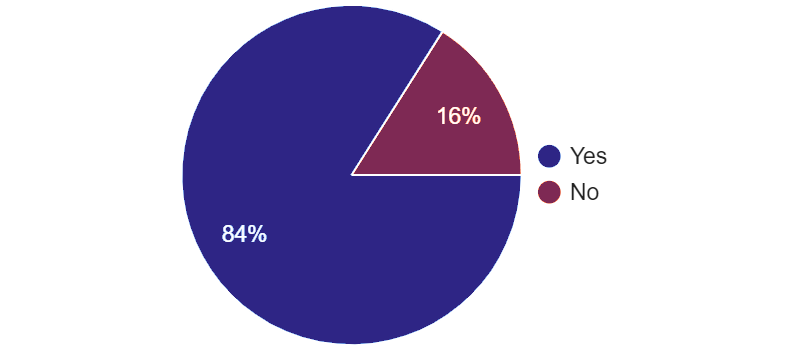}
    \caption{Respondents who have faced added challenges due to COVID-19 or other recent breaches} %additional
    \label{Figure 5}
\end{figure}

\textit{Additional Challenges and Context}:

84\% of respondents acknowledged facing additional challenges due to the pandemic and recent cybersecurity breaches (RQ1) (see Fig. \ref{Figure 5}).

\begin{figure}[ht!]
    \centering
    \includegraphics[width=1\linewidth]{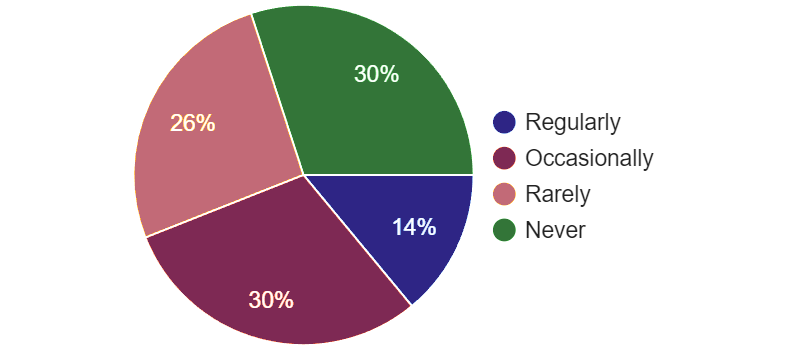}
    \caption{Respondents who get a work call during vacation}
    \label{Figure 6}
\end{figure}

\textit{Vacation Interruptions}: 
Approximately one-third (30\%) of cybersecurity professionals reported receiving work-related calls during planned vacations (see Fig. \ref{Figure 6}). Interestingly, another third (30\%) stated they never receive such calls, while a similar proportion (26\%) reported rare interruptions during their time off.

\subsection{Prevalence of Work Stress and Burnout}

\begin{figure}[ht!]
    \centering
    \includegraphics[width=1\linewidth]{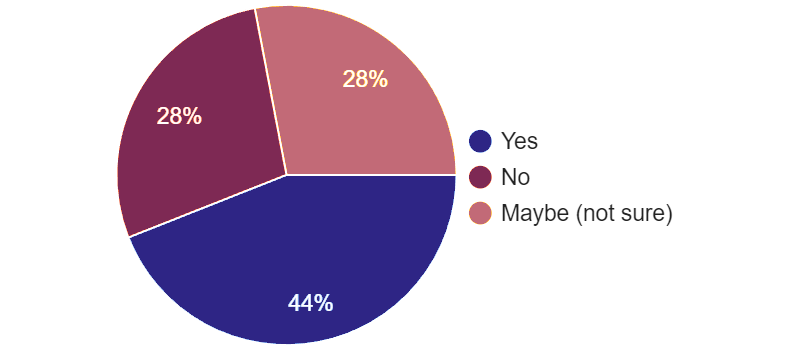}
    \caption{Respondents who are facing or have faced work stress and burnout}
    \label{Figure 7}
\end{figure}

\textit{Stress and Burnout Incidence (RQ1)}: 
Our research highlighted that 44\% of cybersecurity professionals experience work-related stress and burnout (see Fig. \ref{Figure 7}). Meanwhile, 28\% remained uncertain about their stress levels. Only 28\% reported no stress. According to the JD-R model, this outcome is expected in the presence of lower job resources or substantial job demands.

\begin{figure}[ht!]
    \centering
    \includegraphics[width=1\linewidth]{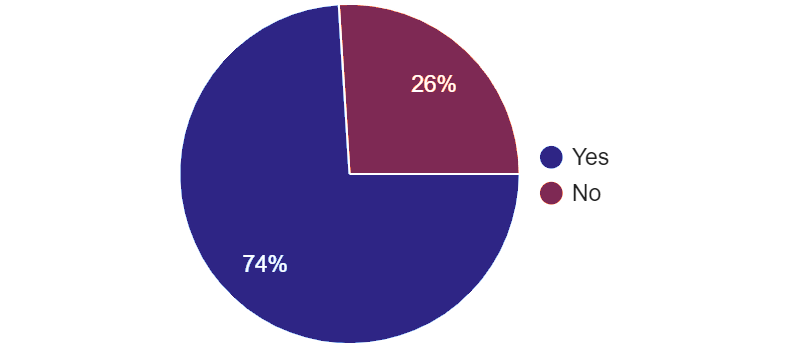}
    \caption{Respondents who know a colleague or co-worker with work stress or burnout}
    \label{Figure 8}
\end{figure}

\textit{Colleague Awareness}: 
Nearly three-quarters (74\%) of survey respondents personally knew a colleague or co-worker who had encountered stress or burnout in their cybersecurity roles (see Fig. \ref{Figure 8}), further emphasizing the prevalence of these issues within the cybersecurity community\footnote{It is understandable that the percentage of respondents experiencing stress or burnout (44\%) differs from the percentage of those aware of others facing such challenges (74\%). This discrepancy is analogous to how the number of movie stars might differ from %those aware of movie stars.
the number of people who are aware of movie stars.}. 
\subsection{Impact and Contributing Factors}
\textit{Underlying Reasons (RQ1)}: 
Our study identified two key factors contributing to stress and burnout, which can be understood through the lens of the JD-R model (see Table \ref{table:reason_comparison}):
\begin{itemize}
    \item \textit{Nature of Cybersecurity Work}: The inherently high-pressure nature of cybersecurity tasks (52\%), representing significant job demands.
    \item \textit{Work Culture and Expectations}: Poor organizational culture (46\%) and unrealistic expectations from management (38\%), suggesting a lack of adequate job resources.
\end{itemize}

\begin{table}[h]
\centering
\caption{Reasons for cybersecurity professionals’ work stress and burnout\tablefootnote{Percentages do not sum to 100 as respondents could select multiple choices.}}
\label{table:reason_comparison}
\resizebox{\columnwidth}{!}{%
\begin{tabular}{|l|c|}
\hline
Reason & Percentage \\ \hline
Nature of cybersecurity work	&  52\% \\ \hline
Poor work culture (including but not limited to corporate  & 46\% \\
politics, work ethics, collaboration, disrespect, etc.)	&  \\ \hline
Unrealistic management expectation	&  38\% \\ \hline
Lack of funding for cybersecurity projects or initiatives	&  36\% \\ \hline
Lack of required knowledge and skills in the team	&  30\% \\ \hline
Lack of communication from cybersecurity management	&  20\% \\ \hline
Unsupported management	&  20\% \\ \hline
\end{tabular}
}
\end{table}

\begin{figure}[ht!]
    \centering
    \includegraphics[width=1\linewidth]{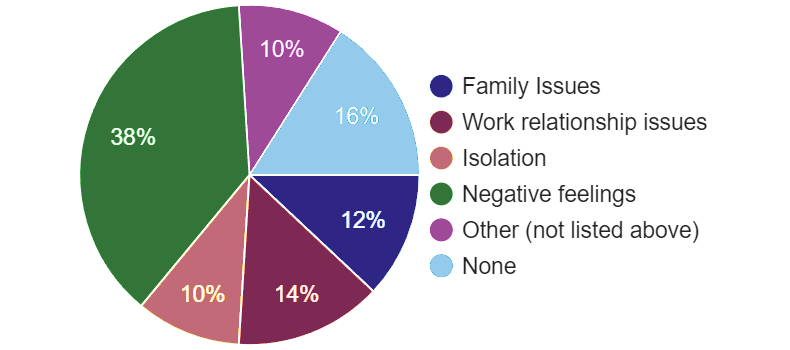}
    \caption{Impacts of work stress and burnout reported by respondents
    }
    \label{Figure 10}
\end{figure}

\textit{Negative Impacts (RQ2)}: 
The primary consequences of stress and burnout in cybersecurity jobs were negative emotions and strained work relationships (see Fig. \ref{Figure 10}), aligning with the JD-R model's prediction of the detrimental effects of high job demands on employee well-being.

\subsection{Reporting and Coping Strategies}

\begin{figure}[ht!]
    \centering
    \includegraphics[width=1\linewidth]{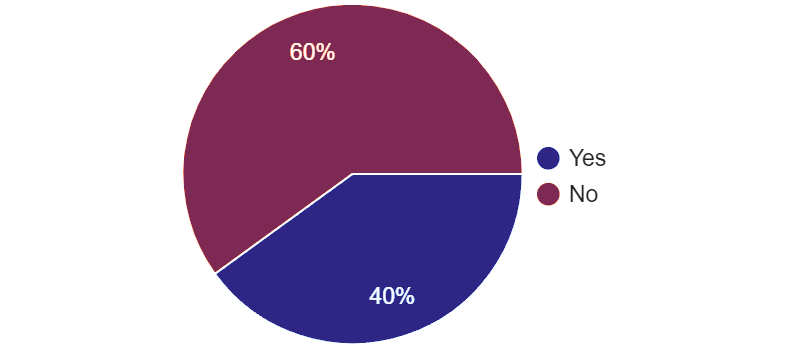}
    \caption{Respondents who reported work stress to management}
    \label{Figure 11}
\end{figure}

\textit{Reporting Behavior}: Surprisingly, most cybersecurity professionals (60\%) are unlikely to report work-related stress and burnout to their management (see Fig. \ref{Figure 11}), indicating a potential lack of job resources, such as organizational support and trust.

\textit{Coping Mechanisms (RQ3)}:
Cybersecurity professionals use a variety of coping mechanism to deal with job related stress (see Table \ref{table:remedy_comparison}):
\begin{itemize}
        \item \textit{Vacations}: Professionals often rely on vacations to alleviate stress (28\%).
        \item \textit{Career Decisions}: Some consider leaving their current organization to mitigate burnout (20\%), while others seek support from family and friends (10\%). 
        
\end{itemize}
These strategies can be interpreted as attempts to manage the high job demands and lack of job resources.

\begin{table}[h]
\centering
\caption{Remedy actions taken by cybersecurity professionals}
\label{table:remedy_comparison}
\resizebox{\columnwidth}{!}{%
\begin{tabular}{|l|c|}
\hline
Support & Percentage \\ \hline
Took vacation or time off                          &                    28\% \\ \hline
Quit the job and moved to another company          &                    20\% \\ \hline
I was not sure what to do                          &                     12\% \\ \hline
Took help from family and friends                  &                     10\% \\ \hline
Did not face stress, burnout and mental health issues &                     10\% \\ \hline
External counseling                                &                     8\% \\ \hline
Changed the role or job within the same company    &                     8\% \\ \hline
Asked for pay raise or promotion                   &                     4\% \\ \hline
\end{tabular}
}
\end{table}

\textit{Preferred Support Strategies (RQ3)}:
Respondents favored the following support approaches to manage stress and burnout (see Table \ref{table:support_comparison}):
\begin{itemize}
    \item Flexible work-life balance such as work-hour flexibility and work from home (40\%)
    \item More vacation and time off (18\%)
    \item Replacing the staff, leader, or manager that causes stress and poor mental health issues (14\%)
    \item Pay increase (12\%)
\end{itemize}

\begin{table}[h]
\centering
\caption{Types of support needed by respondents to improve work stress and burnout}
\label{table:support_comparison}
\resizebox{\columnwidth}{!}{%
\begin{tabular}{|l|c|}
\hline
Support & Percentage \\ \hline
Work-life balance such as work-hour flexibility &                    40\% \\ \hline
More vacation or time off                       &                     18\% \\ \hline
Replacing the staff, leader or manager that causes stress &                     14\% \\ \hline
Pay increase                                    &                     12\% \\ \hline
Resources and tools to manage stress and burnout&                     8\% \\ \hline
New or revised mental health policies and training &                     6\% \\ \hline
Promotion                                       &                     2\% \\ \hline
\end{tabular}
}
\end{table}

\subsection{Correlations and Comparative Analysis}

Within respondents experiencing burnout, 50\% are considering job changes, while for those reporting no burnout, only 7\% of respondents are considering changes. Looking at organization size, burnout was reported by 55\% of respondents from larger organizations (more than 1,000 employees) and by 57\% of respondents from smaller organizations (fewer than 1,000 employees). Thus, organization size by itself doesn't appear to be a predictor of burnout in this data.

Across the survey responses, 90\% of the respondents reported experiencing some form of high job demand. This includes receiving calls during planned vacations, facing unrealistic management expectations, or being stressed due to the nature of cybersecurity work. 46\% of the respondents reported high job demands across multiple indicators. This suggests that while high job demands are widely recognized, nearly half of the respondents dealt with multiple contributing factors.

The factors most positively correlated with burnout were those reporting poor work culture (0.395), those who felt that cybersecurity work is more stressful compared to other IT fields (0.381), and those who noted a lack of organizational support (0.376). The factor most negatively correlated with burnout were respondents who self reported as mid-level professionals (-0.310).

\definecolor{lgreen}{rgb}{0.78, 0.99, 0.78} %lightgreen

\begin{table}[h]
\centering
\caption{Comparison of Burnout Status Between Respondent Job Demands and Job Resources}
\label{table:burnout_comparison}
\resizebox{0.75\columnwidth}{!}{%
\begin{tabular}{|l|c|c|c|c|}
\hline
\textbf{Burnout Status} & \multicolumn{2}{c|}{\textbf{Job Demand}} & \multicolumn{2}{c|}{\textbf{Job Resources}} \\ \cline{2-5} 
                                         & High    & Low & High    & Low   \\ \hline
Yes                                      & \cellcolor{lgreen}47\%           & 20\%   & 31\%           & \cellcolor{lgreen}58\%       \\ \hline
No                                       & 22\%           & 80\%   & 31\%           & 25\%       \\ \hline
Maybe (not sure)                         & 31\%          & 0\%     & 38\%           & 17\%      \\ \hline
\end{tabular}
}
\end{table}

Taking a deeper look, Table \ref{table:burnout_comparison} compares burnout versus job demand and job resources. As predicted by the JD-R model, those self-reporting high job demands and low job resources were the most likely to report burnout, at 47\% and 58\%, respectively.

\section{Discussion}\label{discussion}

The study reveals clear patterns of stress and burnout within the cybersecurity workforce, with only 28\% of respondents reporting an absence of such issues. This finding mirrors previous findings \parencite{Forrester}. To add to the complexity, the COVID-19 pandemic and high-profile cybersecurity breaches have intensified the pressure, creating a climate of heightened anxiety and workload.

The study enhances the theoretical understanding of these challenges by demonstrating the JD-R model’s relevance in cybersecurity, revealing a strong correlation between high job demands, low job resources, and burnout. As predicted by the JD-R model, the data reveals that 47\% of those who reported high job demands also reported experiencing burnout (which is 135\% higher than the 20\% who did not report high job demands). Also, as predicted by the JD-R model, 58\% of those with low job resources also reported experiencing burnout (which is 87.1\% higher than the 31\% reporting high resources). This finding underscores the model's relevance in this context, illustrating that the imbalance created by excessive job demands and insufficient resources is a key predictor of burnout. 

In addition, the data showed that 66\% of respondents perceive cybersecurity roles to be more demanding than other IT positions with respondents further highlighting key job demand variables such as high workload, time pressure, and task complexity. The study also identified a lack of adequate job resources, with many participants pointing to unrealistic expectations from management and insufficient organizational support as critical factors influencing these outcomes. 

Intriguingly, the study indicates that nearly a quarter of respondents expressed uncertainty about the relative stressfulness of cybersecurity compared to other IT jobs. This finding suggests a need for increased awareness and targeted support within the cybersecurity community. Comparative studies across IT sectors could provide valuable insights into stress differentials and inform tailored interventions.

The study extends beyond the workplace, revealing the impact of stress and burnout on cybersecurity professionals' personal lives, with adverse effects on work, family relationships, and overall well-being. The negative impact of neglecting mental health in this workforce highlights the societal consequences of high job demands and inadequate job resources (RQ4), as suggested by the JD-R model. Organizations must recognize that investing in job resources as well as prioritizing employee well-being is not only ethically imperative but also strategically advantageous.

Coping strategies vary, with vacations and social support being common. Some professionals resort to more drastic measures, such as changing jobs or leaving the field altogether—potential talent loss due to stress and burnout warrants pressing attention. The JD-R model suggests that retaining skilled professionals requires proactive interventions and organizational commitment.

The study identifies the reluctance of cybersecurity professionals to report stress and burnout to management as a significant barrier. This silence may stem from several factors, including fear of stigma, lack of trust in leadership, and perceived inadequacy of available support mechanisms. However, this reticence hinders organizations from fully comprehending the extent of the problem and implementing effective solutions (RQ6). The JD-R model suggests that additional job resources could ameliorate this issue. 

To address this critical issue, the study proposes the following recommendations for organizations committed to supporting their cybersecurity workforce (RQ7):

\begin{enumerate}
    \item \textit{Create supportive work environments}: Prioritize flexible work arrangements~\parencite{shifrin2022flexible}, set achievable goals and expectations, and offer competitive compensation packages with adequate benefits, thereby enhancing job resources.
    \item \textit{Implement mindfulness and stress management programs}: To help professionals better cope with high job demands, introduce mindfulness practices to enhance self-awareness, emotional regulation, and stress reduction~\parencite{luken2016systematic}. Conduct workshops on stress management techniques.
    \item \textit{Address understaffing and training needs}: Recognize that an understaffed cybersecurity team can face heightened job demands and health consequences~\parencite{westerlund2004organizational}. Allocate resources strategically and provide the necessary tools, technologies, and training.
\end{enumerate}
These recommendations echo some of the suggestions provided by \textcite{almanza2023} and \textcite{townsend2024}, and align with the JD-R model's emphasis on balancing job demands with job resources to promote employee well-being.

\textcite{brassey2024} argues that by prioritizing employee health, companies can unlock trillions of dollars in economic value. Thus, this study expects that organizations which prioritize the well-being of their cybersecurity workforce could reap long-term benefits. By fostering a supportive environment, implementing evidence-based programs, and addressing systemic challenges, the study anticipates that stress and burnout can be mitigated, ensuring a resilient and effective cybersecurity workforce. The study contributes to the growing discourse on cybersecurity workforce well-being in the digital age, prioritizing mental health for sustained resilience and excellence.

\section{Limitations \& Future Work}\label{futurework}

The sample size within this study is relatively small, and some segments were underrepresented, such as junior level professionals who made up 2\% of the respondent population. However, we feel that the patterns evident in the study provide valuable insights into the mental health and stress levels of cybersecurity professionals.

Participants may not always provide candid answers due to social desirability bias or other factors. As such, future research should explore broader participant pools and delve deeper into the nuanced experiences of professionals in this critical domain.

Additional avenues for further work include the following:

\begin{enumerate}
    \item \textit{Theoretical Foundations}: The JD-R model is well-suited for understanding the stress and burnout issues central to this study. Investigating alternate psychological or sociological theories could provide additional insights. 
    \item \textit{Deeper correlation studies}: The current study identified initial relationships between variables relative to burnout. Further research will explore and discuss relationships across all variables in a greater detail, including potential multivariable interactions, such that the complexity of the issue is more broadly analyzed and that any related conclusions can be discussed in depth.
    
    \item \textit{Longitudinal investigations}: 
    Conduct comprehensive longitudinal studies to track the trajectories of stress and burnout over longer periods. Such research will reveal the cumulative effects as well as potential resilience factors.

\item \textit{Organizational determinants}: 
Investigate the organizational factors that contribute to stress and burnout. Analyze workload allocation, leadership support, team dynamics, and organizational culture. Identifying modifiable factors will inform targeted interventions.

\item \textit{Evaluating intervention strategies}: 
Thoroughly evaluate the efficacy of interventions intended to mitigate stress and burnout. Consider mindfulness programs, workload adjustments, and peer support networks. Comparative analyses across industries will enhance generalizability and provide insights on how to optimize job resources to address job demands.

\item \textit{Cross-industry comparisons}: 
Extend our inquiry beyond the cybersecurity domain. Compare stress and burnout experiences in IT departments with those in other sectors. Understanding sector-specific variations will inform context-sensitive strategies.
\end{enumerate}

\section{Conclusion}\label{conclusion}

This study sheds light on the issue of work-related stress and burnout among cybersecurity professionals, and contributes to a theoretical understanding of occupational stress models through the application of the JD-R framework. Out of 50 surveyed professionals, 44\% reported experiencing these challenges, while an additional 28\% remained uncertain about their conditions. For the well-being of this critical workforce, these findings underscore the significance of addressing the high job demands and the resulting stress and burnout they face, as predicted by the JD-R model.

The demanding nature of cybersecurity roles, coupled with unrealistic expectations and inadequate organizational support, emerged as pivotal job demands contributing to stress and burnout. Aligned with the JD-R model, these heightened job demands can lead to negative outcomes, such as the adverse impacts on personal lives and overall well-being revealed by this study.

Coping strategies, such as seeking vacations and considering career changes, can be interpreted as attempts by cybersecurity professionals to manage the high job demands and lack of job resources. The study's identification of the reluctance to report stress and burnout indicates a potential lack of organizational support and trust, suggesting a deficiency in job resources.

To mitigate these issues and cultivate a more resilient and effective cybersecurity workforce, proactive organizational efforts to address job demands and resources are essential. The study proposes that organizations create supportive work environments, implement mindfulness and stress management programs, and address challenges such as understaffing, unrealistic expectations and training needs. By taking such measures, organizations have the potential to not only improve the mental health of cybersecurity professionals but also strengthen their overall security posture in the face of an increasingly complex digital threat landscape.

\renewcommand*{\bibfont}{\normalfont\small}
\setlength{\bibhang}{15pt}% the hanging indent

\printbibliography
\end{document}